\documentclass[11pt]{article}
\usepackage{moriond,epsfig}
\usepackage{amssymb}

\pdfoutput=1

\def\Journal#1#2#3#4{{#1} {\bf #2}, #3 (#4)}

\def\PLB{{\em Phys. Lett.}  B}
\def\PRL{\em Phys. Rev. Lett.}
\def\PRD{{\em Phys. Rev.} D}

\def\be{\begin{equation}}
\def\ee{\end{equation}}
\def\bea{\begin{eqnarray}}
\def\eea{\end{eqnarray}}

\begin{document}
\vspace*{4cm}
\title{Observation of Diffractively Produced W Bosons in pp Collisions in the CMS Experiment}

\author{ J\"urg Eugster }

\address{on behalf of the CMS Collaboration\\ \vspace{0.2cm} Eidgen\"ossische
  Technische Hochschule (ETH) Z\"urich, Institute for Particle Physics (IPP), Schafmattstrasse 20, 8093
  Z\"urich, Switzerland}

\maketitle\abstracts{A study of forward energy flow in
  leptonically decaying W bosons using data of an
  integrated luminosity corresponding to $36 \; \mathrm{pb^{-1}}$ of pp collisions
  at a center-of-mass energy of $7 \; \mathrm{TeV}$ is presented. This data was recorded with the
  CMS detector during the 2010 running of the LHC. In this sample of W
  events, about 300 events with no significant
  energy deposits in one of the forward calorimeters are observed. This
  corresponds to a large pseudorapidity gap of at least 1.9 units. The
  majority of the charged leptons from these W decays are found in the
  hemisphere opposite to the gap. This gives a strong indication of a
  diffractive component in the W production, which can be explained in
  terms of diffractive parton distribution functions (PDF) which have, on average, a smaller $x$ than the
  conventional parton PDFs.
}

\section{Introduction}
\begin{figure}[t]
\centering
\psfig{figure=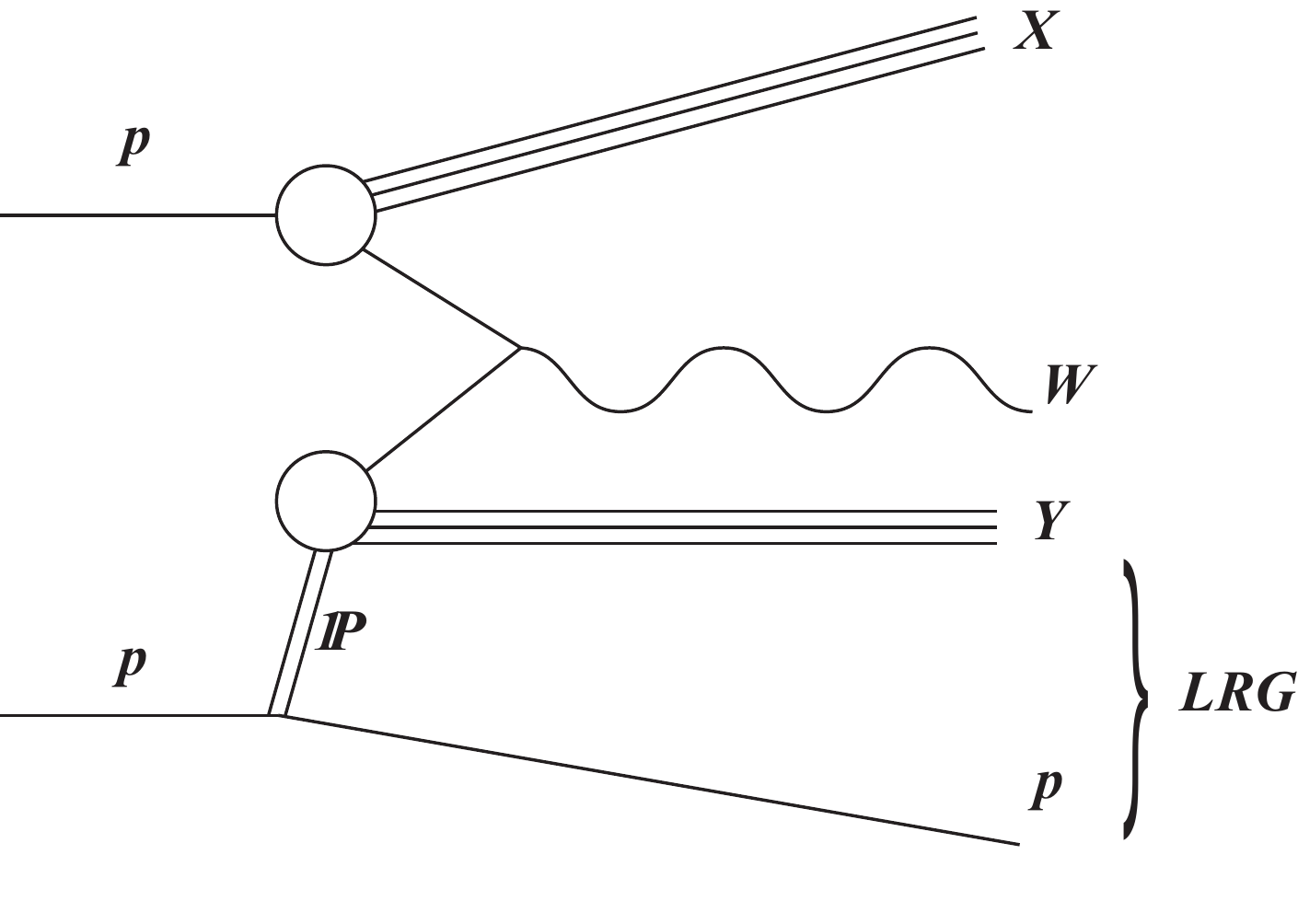,height=5cm} \hspace{2cm}
\psfig{figure=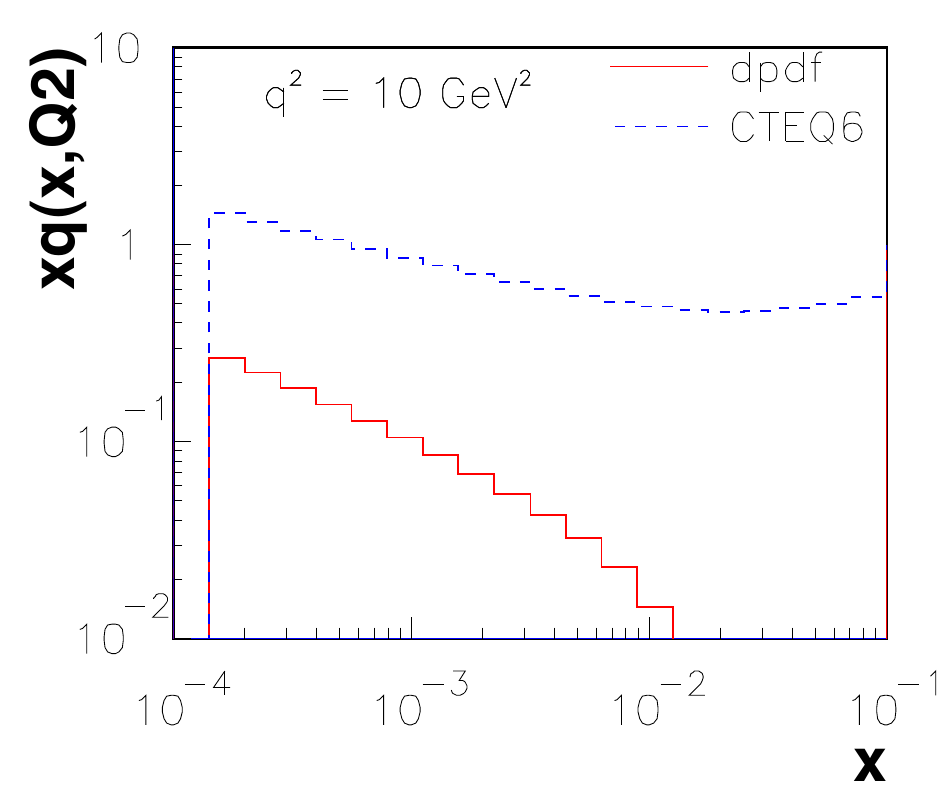,height=5cm}
\caption{(left): Diffractive W production with the exchange of vacuum quantum
  numbers ($\mathbb{P}$). LRG denotes the large
  pseudorapidity gap. (right): Diffractive (solid red line) and
  conventional (dashed blue line) parton
  distribution functions. The cut off for very low $x$ is artificial.
\label{fig:diff}}
\end{figure}

In proton-proton (pp) collisions a significant fraction of the interactions
is expected to arise from single-diffractive (SD) reactions, where one
of the colliding protons emerges intact from the interaction, having
lost only a few percent of its energy. Such SD events may be ascribed
to the exchange of vacuum quantum numbers (often called Pomeron
exchange), interrupting the color flow and, as a consequence
leading to the absence of hadron production over a wide
range of rapidity adjacent to the outgoing proton
direction. Experimentally, these large rapidity gaps will appear as
regions of pseudorapidity, devoid of detectable particles (further
called Large Rapidity Gap: LRG). See Figure~\ref{fig:diff} for
an illustration of diffractive W production.\\

Besides soft-diffractive interactions, hard-diffractive events,
where the LRG signature is found in association with
jets, heavy flavors or W/Z bosons,
have been observed at previous colliders like SPS, HERA and the
Tevatron~\cite{cdf97,ua898,d003,h106,h107,zeus10}. The diffractive
parton distribution functions (PDF) have been introduced and
measured in electron-proton
collisions. In hadron-hadron diffractive interactions, soft multi-parton
interactions occur between the proton remnants, filling the large rapidity gap and reducing the
observed yields of hard-diffractive events by some factor; the so-called
gap survival probability. A recent study by CDF~\cite{cdf10}
indicates that the fractions of diffractively produced W and Z bosons
are $(1.00 \pm 0.11)\%$ and $(0.88 \pm 0.22)\%$ respectively. \\

This summary, which focuses on the observation of diffractively
produced W bosons and follows closely the more extensive analysis presented in
Ref.~\cite{cms12}, where the forward energy flow and central
charged-particle multiplicity and their correlations in leptonically
decaying W and Z boson events are studied in details. It is found that the currently available
Monte Carlo simulations of the underlying event structure, using
non-diffractive soft-hadron production models, do not
describe simultaneously these different  observables.

\section{Event Selection}
A detailed description of the CMS detector can be found
elsewhere~\cite{cms08}. The selection of W events is based on the
presence of exactly one central (i.e. $|\eta| < 1.4$) and isolated
lepton (electron or muon) with high transverse momentum, $p_{T}>25\;\mathrm{GeV}$. In
addition, a large missing transverse momentum of more than $30 \;
\mathrm{GeV}$, from the escaping neutrino, and a transverse mass of
more than $60\;\mathrm{GeV}$ are required. This selection results in an
essentially background free (less than 1\%) W sample.\\

At the LHC, several simultaneous pp interactions can happen in the
same bunch crossing in addition to the selected W event; the so called
pileup. As the LHC instantaneous luminosity was increasing during
2010 operation, the number of such pileup events per bunch crossing was
increasing with time. While the selection efficiency of W events is
independent of the instantaneous luminosity, the
charged-particle multiplicity and the forward energy flow are not, and
thus the LRG signature is strongly affected by pileup; i.e. the gap is
``filled''. In order to limit effects from pileup, events with more
than one vertex are rejected. Pileup can be categorized as hard and soft
events. Hard pileup, producing
detectable central charged particles and thus a vertex, are rejected
by the multi vertex veto. The soft component, presumably from
diffractive proton dissociation, has no detectable transverse
activity in the central region, especially no reconstructed
vertices, but leads to forward energy.\\
This selection results in about $32\,000$ single vertex W events.\\

\section{Forward Energy Flow in W Events}

\begin{figure}[t]
\centering
\psfig{figure=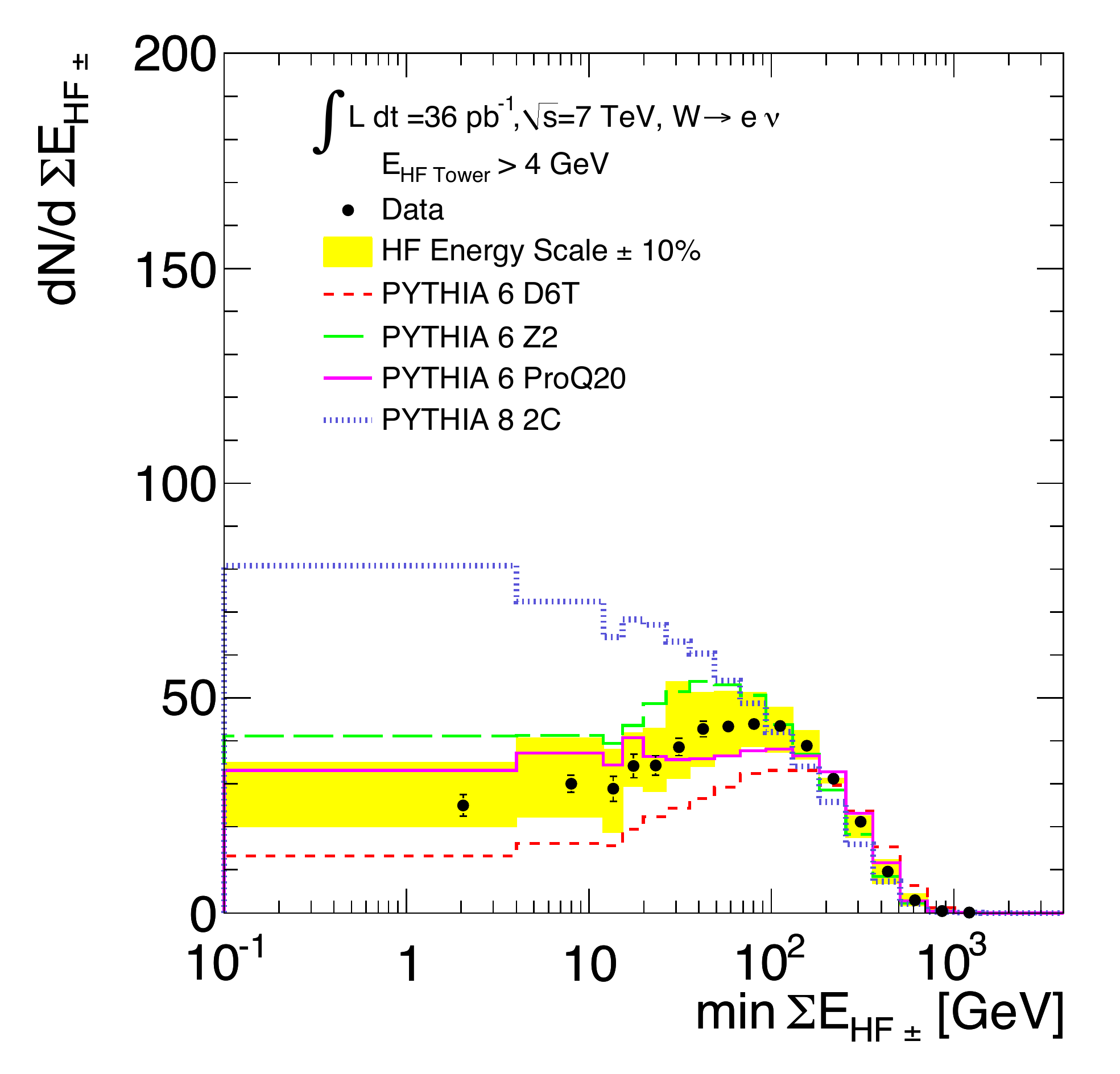,height=6cm} \hspace{0.5cm}
\psfig{figure=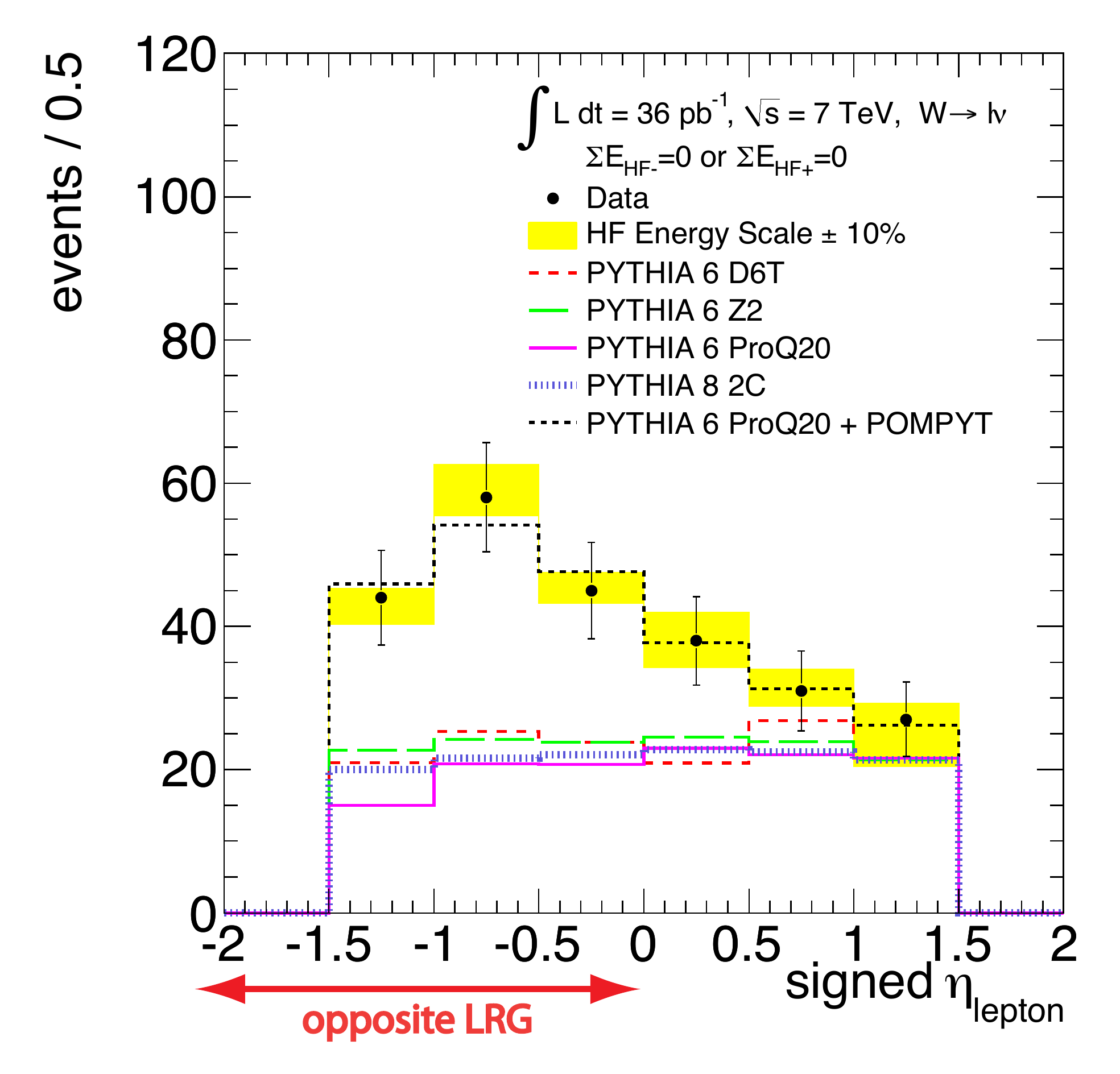,height=6cm}
\caption{(left): minimum of the forward energy deposits in the two hadronic
  forward calorimeters in $\mathrm{W \to e \nu_{e}}$ events. LRG events, having no measurable energy in one
  side of the detector, cluster in the first bin. (right): signed lepton
  pseudorapidity $\eta$. A negative signed $\eta_{lepton}$ means that the
  lepton is in the opposite hemisphere as the gap. The yellow band
  corresponds to the main systematic uncertainty coming from the jet
  energy scale, which is assumed to be 10\%.
\label{fig:fwdE}}
\end{figure}

Figure~\ref{fig:fwdE} (left) shows the minimum of the forward energy deposits in
the two hadronic forward calorimeters (HF+ and HF-, depending on the
$z$-coordinate), which have a pseudorapidity coverage of approximately
$3 < |\eta| < 5$. As already mentioned in the introduction, the
simulation of the forward energy flow in W events is not describing
the data very well. More details can be found in Ref.~\cite{cms12}.\\

A LRG event is defined by the requirement that none of the calorimeter
towers,  in at least one side of the HF, has a measurable energy
deposit above threshold, giving no total energy deposit in one side of the detector. This
results in LRG events with a minimal gap size of $1.9$ units in
$\eta$. This subset is expected to be enhanced by a diffractive
component of W production.\\

Table~\ref{tab:lrg} summarizes the observed LRG event yields and their
fractions to all selected single vertex W events. The dataset has been
split into three periods with different instantaneous luminosities,
and thus different average numbers of pileup events. The
fraction of LRG events decreases with increasing luminosity. As
already mentioned, this can
be explained by HF energy deposits from soft pileup events,
filling the gap.\\

The inefficiency to detect a vertex in events with forward energy from
soft pileup depends on the instantaneous luminosity and is estimated
from zero-bias events. After correcting the observed numbers from
Table~\ref{tab:lrg} for these pileup effects, a constant fraction of
LRG events of $(1.46 \pm 0.09 (\mathrm{stat.}) \pm0.38
(\mathrm{syst.}))\%$, for electrons and muons combined, is found.

\begin{table}[b]
\caption{Number of LRG events with a single vertex and their relative
  fraction to all selected single vertex W events (in brackets), for the total dataset and divided
  into three different periods of different instantaneous luminosities,
  resulting in a different number of average pileup (PU) events.\label{tab:lrg}}
\vspace{0.4cm}
\begin{center}
\begin{tabular}{c||c|c}
avg. number PU events & $\mathrm{W \to e\nu_{e}}$ & $\mathrm{W \to
  \mu\nu_{\mu}}$ \\
\hline
total dataset & 100 (0.71\%) & 145 (0.81\%) \\
$<1$ & 17 (1.13\%) & 31 (1.61\%)  \\
$1-2$ & 57 (0.72\%) & 91 (0.86\%) \\
$>2$ & 26 (0.57\%) & 23 (0.42\%) \\
\end{tabular}
\end{center}
\end{table}

\section{Hemisphere Correlation between LRGs and W bosons}
Figure~\ref{fig:fwdE} (right) shows the distribution of the signed
charged lepton pseudorapidity $\eta_{lepton}$ in W events with a LRG
(electrons and muons combined). The sign is defined to be positive
when the gap and the lepton are in the same hemisphere and negative
otherwise. The data shows that the charged leptons from the W decays
are found more often in the hemisphere opposite to the gap. Defining
an asymmetry as the ratio of the difference between the number of
events in each hemisphere and the sum, the corresponding asymmetry is
$(-21.0 \pm 6.4\%)$. In comparison, the various non-diffractive MC simulations (i.e. PYTHIA
with different tunes) predict a flat signed $\eta_{lepton}$. On the
other hand, events generated with a purely diffractive production
model (i.e. POMPYT), exhibit a strong asymmetry. This can be
explained in terms of diffractive PDFs, which have on average a smaller momentum
fraction $x$ than
the conventional proton PDFs; see Figure~\ref{fig:diff} (right) for an
illustration of this fact. The produced W boson is thus boosted in the
direction of the parton with the larger $x$, which is typically the
direction of the dissociated proton, i.e. opposite to the gap.\\

The signed $\eta_{lepton}$ from non-diffractive and purely diffractive
Monte Carlo is fitted to the distribution from data, resulting in a
fraction of diffractive events of $(50.0 \pm 9.3 (\mathrm{stat.}) \pm
5.2 (\mathrm{syst.}))\%$, assuming the PYTHIA6 Pro-Q20 tune as the
non-diffractive component. The other tunes yield similar
results. Figure~\ref{fig:fwdE} (right) shows this combined simulation and
the purely non-diffractive components for the other tunes.\\

The asymmetry in the signed $\eta_{lepton}$ distribution for non-LRG
events decreases with increasing forward energy deposits; e.g. for HF
energy deposits of $20-100 \; \mathrm{GeV}$, $200-400 \; \mathrm{GeV}$
and $>500 \; \mathrm{GeV}$, the asymmetries are $(-3.5 \pm 1.1)\%$, $(-2.7 \pm 1.0)\%$, and
$(0.9 \pm 2.3)\%$ respectively. The small residual asymmetry in low HF
energy events is insignificant in comparison to the one for LRG
events. However it could be explained by the presence of a diffractive
component in which the LRG signature is destroyed by 
multi-parton interactions or undetected pileup events. For higher HF
energy deposits, the
asymmetry vanishes completely.

\section{Summary}
Out of approximately $32\,000$ W events about $300$ are found with a
LRG. The fraction of events with LRG, as predicted by the
non-diffractive Monte Carlo simulation (PYTHIA) is strongly tune dependent. An asymmetry
between the number of events with the charged lepton in the opposite
and the same hemisphere as the gap is observed. Such an asymmetry is
predicted by the purely diffractive simulation (POMPYT). Using an admixture
of diffractive and non-diffractive Monte Carlo describes the
data. The fraction of the diffractive component is determined from a
binned maximum likelihood fit to be $(50.0 \pm 9.3 (\mathrm{stat.}) \pm
5.2 (\mathrm{syst.}))\%$, thus providing strong evidence for
diffractive W production at the LHC.


\end{document}